\documentclass[useAMS,usenatbib]{mn2e}

\usepackage[dvips,draft]{graphicx}

\title[Eclipsing Light Curve for Accretion Flows around Rotating Black Hole]
{Eclipsing Light Curve for Accretion Flow around Rotating Black Hole
and Atmospheric Effects of Companion Star}
\author[Takahashi \& Watarai]{R. Takahashi$^{1}$\thanks{E-mail:
rohta@provence.c.u-tokyo.ac.jp}
and 
K. Watarai$^{2}$
\\
$^{1}$
Department of Earth Science and Astronomy, 
Graduate School of Arts and Sciences, University of Tokyo, Tokyo 153-8902, Japan\\
$^{2}$Astronomical Institute, Osaka Kyoiku University, Kashiwara, Osaka 582-8582
}
\begin{document}

\date{Accepted 200X December XX. Received 200X December XX; in original form 200X August 5}

\pagerange{\pageref{firstpage}--\pageref{lastpage}} \pubyear{200X}

\maketitle

\label{firstpage}

\begin{abstract}
We calculate eclipsing light curves for accretion flows around a rotating black hole taking into account the atmospheric effects of the companion star. In the cases of no atmospheric effects, the light curves contain the information of the black hole spin because most of X-ray photons around 1 keV usually comes from the blue-shifted part of the accretion flow near the black-hole shadow and the size and the position of the black-hole shadow depend on the spin. In these cases, when most of the emission comes from the vicinity of the event horizon, the light curves becomes asymmetric at ingress and egress. We next investigate the atmospheric absorption and scattering effects of the companion stars. By using the solar-type atmospheric model, we have taken into account the atmospheric effects of the companion star, such as the photoionization by HI and HeI. We found that the eclipsing light curves observed at 1 keV possibly contain the information of the black-hole spin. However, in our atmospheric model, the effects of the atmosphere are much larger than the effects of the black-hole spin. Therefore, even in the case that the light-curves contain the information of the black hole spin, it may be difficult to extract the information of the black hole spin if we do not have the realistic atmospheric profiles, such as, the temperature, the number densities for several elements. Even in such cases, the light curve asymmetries due to the rotation of the accretion disc exist. Only when we have the reliable atmospheric model, in principle, the information of the strong-gravity regions, such as, the black hole spin, can be obtained from the eclipsing light curves.
\end{abstract}


\section{Introduction}
Elucidating the nature of the physics of extremely strong gravity around 
black holes is one of the greatest challenges in astrophysics in this century. 
In order to obtain physical information of the strong gravity region in the vicinity of 
black holes, direct imaging of a black hole is an extremely promising method. 
The emission from the vicinity of the event horizon contains information 
about the physical parameters of a black hole. 
Since the apparent shapes of the black hole shadows and the appearances of the 
circular orbit around the black hole 
are deformed by the effects of the black hole's rotation and charge
(e.g. Cunningham \& Bardeen 1972, 1973, Bardeen 1973, Takahashi 2004, 2005, 
Zakharov et al. 2005), 
the size and the position of the most luminous parts generally existing 
in the vicinity of the shadow depend on the physical parameters of black holes. 
In general, 
methods for probing strong-field gravity near supermassive black holes  
or stellar-mass black holes are quite limited. 
Although the past VLBI observations give the images of the galactic centers 
with the spatial resolution within 100 Schwarzschild radii   
(e.g., for Sgr A*, Shen et al. 2005, for M87, Junor, Biretta \& Livio 1999), 
the direct mapping of the shadows of both supermassive and stellar-mass 
black holes has not been performed so far. 
Several future interferometers have plans to image the shadows 
around massive black holes: e.g. VSOP-2 for radio (Hirabayashi et al. 2005), 
MAXIM for soft X-ray (see MAXIM web page: http://maxim.gsfc.nasa.gov/). 
While these planned interferometers will image the massive black hole, 
direct imaging of the stellar-mass black holes are difficult because 
the apparent size of the black hole shadow is extremely small. 
For example, in the case of the black hole with 10 solar mass 
at the distance of 10 pc, 
the apparent size of the black hole shadow ($\sim$ 5 Schwarzschild radii) 
is about 0.1 micro-arcsecond. 
So, even with the future planned radio or X-ray interferometer, 
the direct mapping of the shadow of the stellar-mass black hole 
is practically impossible. 
However, if the stellar-mass black hole in the binary system is occulted 
by its companion star, the eclipsing light curves give 
the direct information for the spatial distribution of the brightness 
patterns of the accretion flows around the black hole. 
This is the primary motivation of the present study. 
In terms of the studies about the strong-gravity region around the 
stellar-mass black hole, 
some observations in the field of X-ray spectroscopy 
proposed the values of the spin parameters of stellar-mass black holes 
(e.g. Zhang et al. 1997, Miller et al. 2002). 
In addition, recently the works in this field 
have achieved the significant advances for understanding the accretion 
flows in the strong-gravity regions around the black holes 
(e.g. Davis et al. 2006, Done \& Gierlinski 2003, 2006, Gierlinski \& Done 2004). 

In the present study, we study the method to investigate the physical 
information included in the strong-gravity region around 
the black hole by light-curve analysis for an accretion disc around 
a black hole occulted by its companion star. 
The basic idea for the eclipsing light curve analysis 
was first proposed by Fukue (1987). 
Watarai, Takahashi \& Fukue (2005) investigated the eclipsing light curves 
for the supercritical accretion flows which may be suggested 
as the accretion flow model for some X-ray black-hole binaries.  
In principle, 
light curves obtained at the time of eclipsing have information about 
the black hole if the region around the black hole is optically thin.  
However, the real observational data of the eclipsing light curves 
may contain the additional effects which are not sometimes simple. 
One of such effects is the atmospheric effects of the companion star. 
Recently, Pietsch et al. (2006) give the observational data of the 
eclipsing light curves for the X-ray binary M33 X-7 
detected by {\it Chandra}. 
In soft X-ray, the atmospheric effects in the companion star may vanish the 
relativistic effects in the eclipsing light curves. 
In other words, the eclipsing light curves also contain the information of 
the atmosphere of the companion star. 
Thus, the eclipsing light curve analysis will be used for 
the investigation of the atmosphere of the companion star. 
In the present study,  
we also investigate the companion star's atmospheric effects in the 
eclipsing light curves of the black-hole binaries, 
and compare these effects with the effects of the black hole's rotation.  
In \S 2, we give the calculation method of the eclipsing light curves. 
The adopted accretion disk model, the assumed spectral model and 
the calculation method of the eclipsing light curves without the atmospheric 
effects of the companion star are given in \S 2.1. 
The assumed atmospheric model of the companion star and 
the calculation method of the eclipsing light curves with the atmospheric 
effects of the companion star are given in \S 2.2. 
In \S 3, we calculate eclipsing light curves 
without the atmospheric effects of the companion star. 
After we calculate the eclipsing light curves around rotating black holes 
in \S 3.1, 
the analysis by using the normalized light curves and 
the statistical quantities of the skewness and the kurtosis are 
performed in \S 3.2 and \S 3.3, respectively. 
In \S 4, we calculate eclipsing light curves 
with the atmospheric effects of the companion star. 
After we show the sample images of the black hole shadows 
with the atmospheric effects in \S 4.1
and the normalized light curves in \S 4.2, 
the eclipsing light curves are analysed with respect to 
the observed photon energy in \S 4.3, 
the black hole spin in \S 4.4, 
and 
the inclination angle between the rotation axis of the disc and the direction 
of the observer in \S 4.5.   
Related topics are discussed in \S 5. 
We give concluding remarks in the last section, \S 6. 

\section{Calculation Method for Eclipsing Light Curves}
\subsection{Eclipsing Light Curves without Atmospheric Effects}

\begin{figure}
\includegraphics[width=80mm]{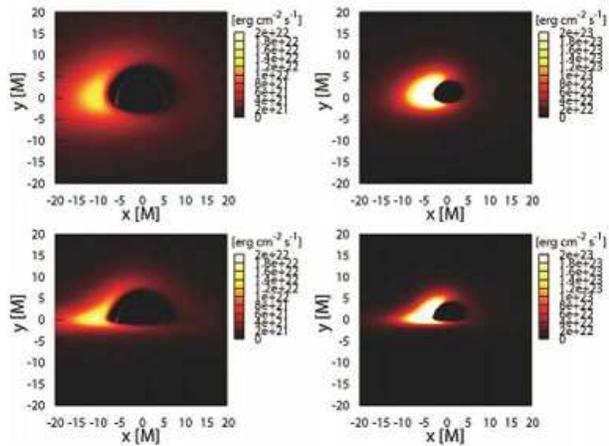}
\caption{
Bolometric flux distribution (erg cm$^{-2}$ s$^{-1}$) 
for black-hole spins of $a/M=0$ ({\it left panels}) and $1$ 
({\it right panels}), and inclination angles of $i=60^\circ$ 
({\it top panels}) and $80^\circ$ ({\it bottom panels}).  
The relativistic standard discs are adopted. 
The inner edge of the accretion disc is assumed to be the marginally 
stable circular orbit. 
For non-rotating black holes, indirect images can be seen in the region 
within the inner stable circular orbit. 
The flux distributions are asymmetric due to the disc rotation and the 
frame-dragging effects. 
The shapes of the shadows casted by the black holes are asymmetric due to 
frame-dragging effects around the rotating black holes. 
}
\label{fig:BHimage}
\end{figure}
%

\begin{figure}
\includegraphics[width=80mm]{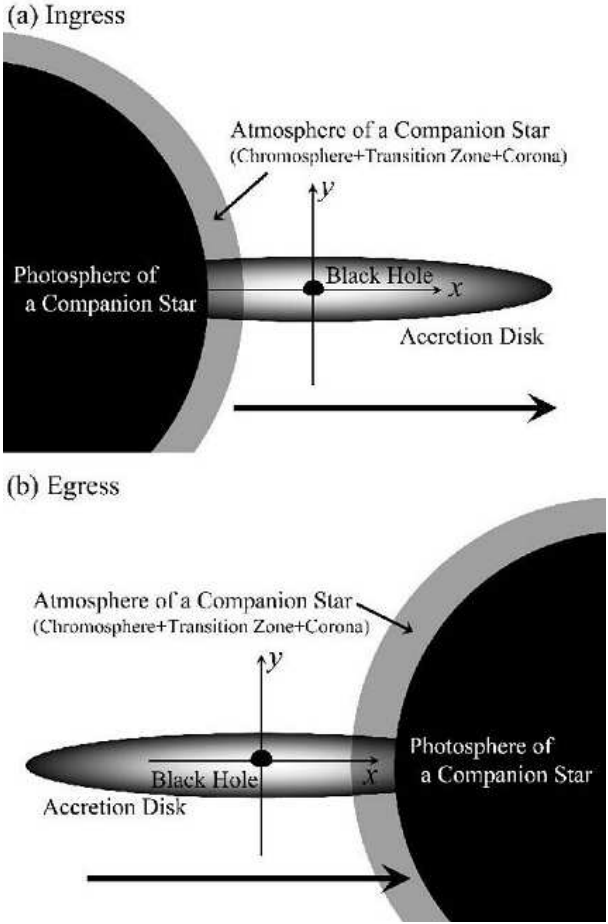}
\caption{
Schematic diagrams for the eclipsing black-hole binaries when the ingress ({\it top panel}) and 
the egress ({\it bottom panel}). 
}
\label{fig:ingress_egress}
\end{figure}

In order to calculate a light curve during an eclipse, 
some assumptions were required. 
In the present study, we assume the relativistic standard disc 
(Novikov \& Thorne 1973, Page \& Thorne 1974) at equatorial plane. 
Based on this model, 
the radial profile of the effective temperature of the accretion flow 
can be calculated.  
Then, at each radius, we assume the black body spectrum with the effective 
temperature calculated from the relativistic standard accretion disc model. 
The calculation method for the observed spectrum of the emission 
in the Kerr metric is established (e.g. Cunningham \& Bardeen 1973). 
At each light ray, 
the observed 
specific intensity, $I_{\nu_{\rm obs}}$ [erg s$^{-1}$ cm$^{-2}$ str$^{-1}$ Hz$^{-1}$], 
of the accretion disc observed at the photon frequency $\nu_{\rm obs}$ is calculated as 
%
$I_{\nu_{\rm obs}}=g^3 I_{\nu_{\rm rest}}$,
%
where $\nu_{\rm rest}$ is the photon frequency measured in the local rest 
frame of the accretion flow, $T_{\rm eff}$ is the temperature 
of the relativistic standard accretion disc at the radius where the light ray intersects, 
$I_{\nu_{\rm rest}}$ is the specific intensity measured at the local rest frame 
of the accretion disc, and $g$ is the frequency ratio defined 
$g=\nu_{\rm obs}/\nu_{\rm rest}$. 
The frequency ratio $g$ is calculated same as Cunningham \& Bardeen (1973). 
In this study, we assume $I_{\nu_{\rm rest}}$ is the black body spectrum with 
the effective temperature at the radius where the light ray intersects, i.e. 
$I_{\nu_{\rm rest}}=B_{\nu_{\rm rest}}(T)$, where $T$ is the temperature of the 
accretion disc.   
By using the relation $d\nu_{\rm obs}=gd\nu_{\rm rest}$ and the assumption of the 
isotropic radiation at the local rest frame of the accretion disc, 
the observed flux $F_{\rm obs}$ [erg s$^{-1}$ cm$^{-2}$] 
is calculated as $F_{\rm obs}=(1/d^2) \int g I_{\nu_{\rm obs}}d\nu_{\rm obs}
=(1/d^2)\int g^4 I_{\nu_{\rm rest}} d\nu_{\rm rest}$ where $d$ is the distance 
to the accretion disc.  
By integrating the observed flux in $x$-$y$ plane, we can calculate the observed 
bolometric luminosity $L_{\rm bol}$ [erg s$^{-1}$]. 
On the other hand, the observed flux at the observed photon frequency 
$\nu_{\rm obs}$, 
$F^{\rm obs}_{\nu_{\rm obs}}$ [erg s$^{-1}$ cm$^{-2}$ Hz$^{-1}$], 
is calculated by using the delta-function as 
$F_{\nu_{\rm obs}}=(1/d^2)\int g^4 I_{\nu_{\rm rest}} 
\delta (\nu_{\rm obs}-g\nu_{\rm rest}) d\nu_{\rm rest}$. 
By integrating the flux in $x$-$y$ plane, the luminosity at the observed frequency 
$\nu_{\rm obs}$, $L_{\nu_{\rm obs}}$ [erg s$^{-1}$ Hz$^{-1}$] is calculated. 
The luminosity are calculated by the ray-tracing method, 
which is commonly used in the past studies (e.g. Fukue \& Yokoyama 1988, 
Takahashi 2004). 
When we transform the luminosity from the local rest frame to 
the Boyer-Lindquist frame by calculations of tetrads, 
the relativistic effects are included, that is, 
the special relativistic effects such as beaming effects and Doppler 
boosting, and the general relativistic effects such as 
gravitational redshift, gravitational lensing and frame-dragging effects. 
The frequency ratio $g$ includes these special and general relativistic effects. 
For all calculations in the present study, 
we assume a black hole mass of $10M_\odot$ and mass accretion rate of 
$1\dot{M}_{\rm Edd}$, and the disc thickness is neglected.   
Here $\dot{M}_{\rm Edd}$ is the Eddington mass accretion rate defined 
as $\dot{M}_{\rm Edd} \equiv L_{\rm Edd}/c^2 =1.4\times 
10^{17} (M/M_\odot)$ 
[g/s] where $L_{\rm Edd}$ is the Eddington luminosity 
(e.g. Kato, Fukue \& Mineshige 1998). 
Also, we implicitly assume that the binary system has a relatively high 
inclination angle, $i>60^\circ$. 
In figure \ref{fig:BHimage}, we show 
examples of the spatial distribution of the observed flux of the accretion flows 
in the vicinity of the black holes. 
For all panels of figure \ref{fig:BHimage}, 
the mass center of the black hole is at the center of the image. 
The width and the height of the images in figure \ref{fig:BHimage} are 
both 20 $M$ where $M$ is the gravitational radius and 1 $M$ corresponds 
to 0.5 Schwarzschild radius.
We define $x$-axis and $y$-axis as the horizontal line and the vertical line 
crossing at the center of the each image of figure \ref{fig:BHimage}. 
In figure \ref{fig:BHimage}, 
the position of the mass center of the black hole is at $(x,~y)=(0,~0)$, and 
the displayed region is $-10$ [$M$]$<x<10$ [$M$] and $-10$ [$M$]$<y<10$ [$M$]. 
Based on the calculations of the observed luminosity described above, 
we calculate the eclipsing light curves. 
In figure \ref{fig:ingress_egress}, 
we show the schematic diagrams for the eclipsing black-hole binaries 
when the ingress ({\it top panel}) and 
the egress ({\it bottom panel}).
We also plot $x$-axis and $y$-axis in panels of figure \ref{fig:ingress_egress}. 
When the ingress and the egress phases described in figure \ref{fig:ingress_egress}, 
the companion star is crossing the line of sight between the observer and 
the black hole binary along the direction of $x$-axis. 
Since the size of the companion star is generally much larger than the emission region 
around the black hole, 
we assume the carve of the stellar surface as the straight line in our calculations.  
Although in our calculations the brightness profiles of 
the accretion disc are calculated within $r<200$ [$M$], 
the light curves are calculated in the range of $-60$ [$M$]$<x<60$ [$M$]. 
That is, in terms of the range in $y$-axis direction, we use the brightness profile 
of the disc in the range of $-200$ [$M$]$<y<200$ [$M$], but 
the brightness profiles of the accretion disc are given within $r<200$ [$M$]. 
%

\subsection{Atmospheric Model of Companion Star and 
Eclipsing Light Curves with Atmospheric Effects}
\begin{figure}
\includegraphics[width=80mm]{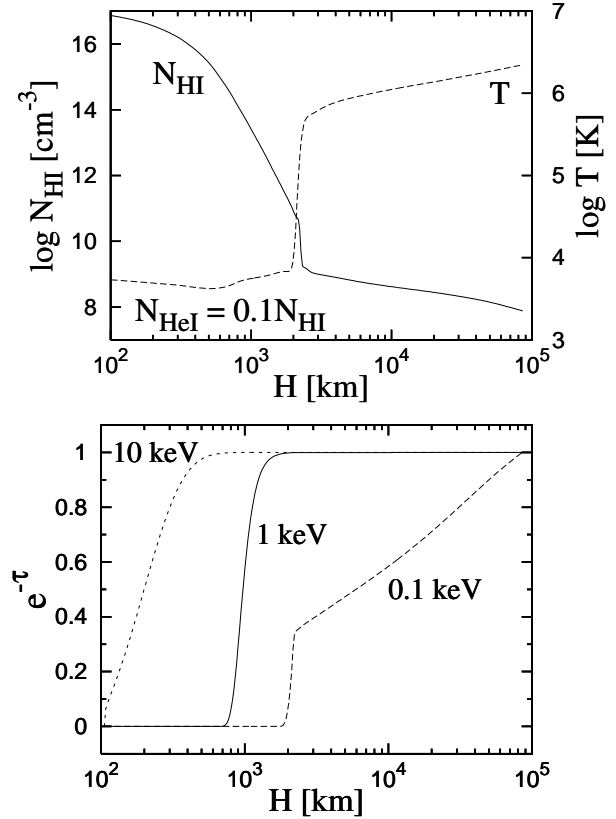}
\caption{
HI number density $N_{\rm HI}$ [cm$^{-3}$] ({\it solid line}) 
and temperature $T$ [K] ({\it dashed line}) of the adopted atmosphere 
model ({\it top panel}). 
The atmosphere model has been constructed for the solar atmosphere 
(Daw et al. 1995).  
The number density of HeI, $N_{\rm HeI}$, is assumed to be 
$N_{\rm HeI}=0.1N_{\rm HI}$. 
The corresponding transparency ratio, $e^{-\tau}$, 
of the modeled atmosphere 
({\it bottom panel}) for X-ray: 
0.1 keV ({\it long dashed line}), 1 keV ({\it solid line}), 
10 keV ({\it short dashed line}). 
Here, $\tau$ is the optical depth of the photoionization absorption 
by HI and HeII.    
}
\label{fig:tau}
\end{figure}

In the present study, we use the atmospheric model calculated for 
the solar atmosphere 
(Daw et al. 1995, Fontenla et al. 1993, Gabriel 1976). 
In the top panel of figure \ref{fig:tau}, we plot 
HI number density $N_{\rm HI}$ [cm$^{-3}$] ({\it solid line}) 
and temperature $T$ [K] ({\it dashed line}) of the adopted atmosphere 
model. 
Here, $H$ [km] denote the height above the photosphere. 
Temperature of the atmosphere suddenly increase and the number density 
decrease in the transition zone between 
the lower chromosphere ($\sim 10^{3-4}$ [K]) and 
the upper corona ($\sim 10^6$ [K]). 
The two separate temperatures of HI correspond to 
two stable regions of the cooling function of HI 
(e.g. chapter 6 in Spitzer 1978).  

\begin{table}
 \centering
  \caption{Cross sections [cm$^2$] for photoionization absorption 
due to HI, HeI, HeII and Compton scattering in X-ray of 0.1 keV, 1 keV and 
10 keV.}
  \begin{tabular}{@{}cccc@{}}
  \hline
   & 0.1 keV & 1 keV & 10 keV \\
 \hline
$\sigma_{\rm HI}$  & $1.9\times 10^{-20}$ 
	& $1.1\times 10^{-23}$ & $4.5\times 10^{-27}$ \\
$\sigma_{\rm HeI}$  & $3.9\times 10^{-19}$ 
	& $3.4\times 10^{-22}$ & $1.4\times 10^{-25}$ \\
$\sigma_{\rm HeII}$  & $3.0\times 10^{-19}$
	& $2.7\times 10^{-22}$ & $1.3\times 10^{-25}$ \\
$\sigma_{\rm Comp}$  & $1.4\times 10^{-25}$ 
	& $1.3\times 10^{-25}$ & $6.4\times 10^{-25}$ \\
\hline
\end{tabular}
\label{table:cs}
\end{table}

\begin{figure}
\includegraphics[width=80mm]{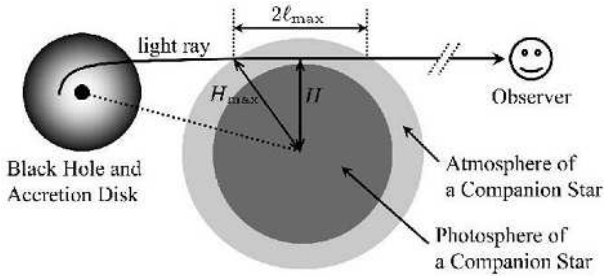}
\caption{
Schematic diagrams for the observer, the eclipsing black-hole binaries, 
the companion star with its atmosphere and the light ray penetrating 
the atmosphere. 
Here, the light ray penetrate the atmosphere with the distance $2\ell_{\rm max}$. 
$H_{\rm max}$ is the distance from the center of the companion star to 
the maximum height of the atmosphere, 
and $H$ is the distance between the center of the companion star 
and the point where the light ray penetrate the atmosphere.  
}
\label{fig:BHRT}
\end{figure}

In figure \ref{fig:BHRT}, 
we show the schematic diagrams for the observer, the eclipsing black-hole binaries, 
the companion star with its atmosphere and the light ray penetrating 
the atmosphere. 
Here, the light ray penetrates the atmosphere with the length of $2\ell_{\rm max}$. 
$H_{\rm max}$ is the length from the center of the companion star to 
the maximum height of the atmosphere, 
and $H$ is the length between the center of the companion star 
and the point where the light ray penetrate the atmosphere.  
The optical depth, $\tau(\nu)$, in the atmosphere can be calculated as 
\begin{equation}
\tau(\nu)
=\int d\ell~(\sigma_{\rm abs}N_{\rm abs}+\sigma_{\rm sca}N_{\rm sca}), 
\end{equation}
where $\sigma_{\rm abs}$ and $\sigma_{\rm sca}$ 
are cross sections of the absorption and scattering, respectively, 
and $N_{\rm abs}$ and $N_{\rm sca}$ are the number densities 
of the absorber and the scattering medium, respectively.  
The integrations are performed along the atmosphere of the companion star 
in the line of sight. 
In soft X-ray band, the photoionization absorption by hydrogen and helium 
contribute to opacity. 
The photoionization absorption cross sections of HI, HeI and HeII 
in soft X-ray are summarized in table \ref{table:cs}. 
We also calculate the Compton scattering cross section 
in the same table. 
Since for 0.1 keV and 1 keV 
the scattering cross section is significantly lower than 
the absorption cross sections, we can neglect the effects of 
the Compton scatterings. 
On the other hand, for 10 keV the scattering effects are dominated. 
In this case, we only take into account the out-going photons 
as the scattering effects for simplicity. 
We see that the photoionization absorption cross sections of HI, HeI 
and HeII are comparable in soft X-ray. 
Calculations by Fontenla et al (1993) shows that 
the number density of HeII, $N_{\rm HeII}$, is significantly 
lower, e.g. by order 4, than that of HI, $N_{\rm HI}$, in the solar 
atmosphere, i.e. $N_{\rm HeII}\ll N_{\rm HI}$, and 
the number density of HeI is lower by about 
order 1 than that of HI through the atmosphere, i.e. 
$N_{\rm HeI}\sim N_{\rm HI}$. 
Therefore, in the present study, 
we neglect the scattering effects in the atmosphere for 0.1 keV and 
the absorption by HeII for 0.1 keV, 1 keV and 10 keV. 
Then, we calculate the optical depth in the atmosphere as 
\begin{equation}
\tau(\nu)=2\int_0^{\ell_{\rm max}} d\ell~
(\sigma_{\rm HI}N_{\rm HI}+\sigma_{\rm HeI}N_{\rm HeI}
+\sigma_{\rm Comp}N_e),  
\end{equation} 
where $\ell_{\rm max}=(H_{\rm max}^2-H^2)^{1/2}$, $H_{\rm max}$ is set to 
$10^5$ [km], the number density of electrons $N_{e}$ is calculated only 
in the photosphere based on numerical table given by 
Fontenla et al. (1993). 
However, 
according to Fontenla et al. (1993) 
the number density of electrons in the atmosphere 
is lower than that of HI by factor 3-4 in the almost region 
of photosphere, then the scattering effects are negligible in 
the photosphere. 
Near the transition zone, 
the number density of the electron is lower than that of HI 
within factor 1, and for cases of 10 keV 
the scattering effects contribute to the optical depth. 
By using the optical depth calculated above, 
we solve the radiative transfer equation along the atmosphere in the 
line of sight. 
The brightness decreases along the light rays in the atmosphere 
by the exponential of the optical depth. 
By using the optical depth $\tau$ calculated above, 
the observed luminosity with the atmospheric effects is calculated 
by multiplying the factor $e^{-\tau(\nu_{\rm obs})}$ to 
the observed luminosity without the effects of the atmospheric effects. 
The observed luminosity without the atmospheric effects corresponds 
to the case of $\tau=0$. 
Since the distance between the accretion disc and the atmosphere of the 
companion star is very large, here 
we evaluate the optical depth $\tau$ at 
the observed photon frequency $\nu_{\rm obs}$. 
In the bottom panel of figure \ref{fig:tau}, we show 
the corresponding transparency ratio, $e^{-\tau}$, 
of the modeled atmosphere for X-ray: 
0.1 keV ({\it long dashed line}), 1 keV ({\it solid line}), 
10 keV ({\it short dashed line}).  
In the cases of 1 keV and 10 keV, 
while the lines in the corona are optically thin, 
the photoionization absorption is effective in the photosphere. 
By using the optical depths calculated as figure \ref{fig:tau}, 
we calculate the eclipsing light curves with the atmospheric effects 
of the companion star in later section. 
%

In the present study, we do not include contributions from the heavy elements 
which are important in some cases because the photoelectric cross sections are 
generally proportional to $Z^4$ where $Z$ is the proton number. 
However, we do not intend to give the quantitatively precise information of the 
eclipsing light curves in this section, but to roughly investigate the possibility 
whether the atmospheric absorption effects can be negligible or not. 
Moreover, even if we calculate more precise light curves including 
the effects of the heavy elements, there is no observational targets having the 
solar-type atmosphere for which the precise light curves can be used to fitting 
the observational data (See also discussion for the case of M33 X-7).

\section{Eclipsing Light Curves without Atmospheric Effects of 
Companion Star}

\subsection{Calculated Light Curves}
\begin{figure}
\includegraphics[width=80mm]{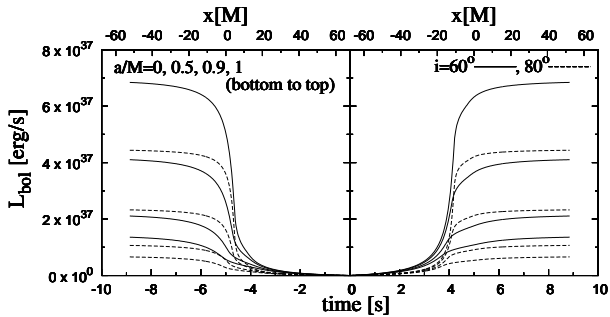}
\caption{
Eclipsed light curves before ({\it left panel}) and after 
({\it right panel}) a total eclipse by a companion stars. 
For each panel, 
the lower abscissa shows the time of eclipse and the upper abscissa 
shows the position of eclipse with respect to the position of 
black hole ($x=0$). 
The inclination angles are $60^\circ$ ({\it solid lines}) and 
$80^\circ$ ({\it dashed lines}). 
The black hole spins are $0$, $0.5$, $0.9$ and $1$ from bottom to top 
for $i=60^\circ$ and $80^\circ$, respectively. 
The lines showing zero time (time$=0$ s) indicate the boundary of a complete eclipse. 
We assumed an orbital velocity of $200$ km s$^{-1}$, a black hole mass of 
$10 M_\odot$ and mass accretion rate of $1 \dot{M}_{\rm Edd}$. 
}
\label{fig:LC}
\end{figure}

%
%
By using the observed bolometric luminosity $L_{\rm bol}$, 
in figure \ref{fig:LC} we show the light curves 
at the ingress ({\it left panel}) and egress ({\it right panel}) 
of an eclipse. 
The observed bolometric luminosity is the function of the position 
of the surface $x_s$ of the companion star, i.e. $L_{\rm bol}=L_{\rm bol}(x_s)$. 
The observed bolometric luminosity is calculated by integrating the observed flux 
in $x$-$y$ plane. 
For the ingress phase with the stellar surface at $x$, the bolometric luminosity 
$L_{\rm bol}(x)$ is given as 
\begin{equation}
L_{\rm bol}(x)
	=\frac{1}{d^2}\int_{x}^{200[M]}dx^\prime\int_{-200[M]}^{200[M]}dy^\prime
	\int d\nu_{\rm rest}~g^4~I_{\nu_{\rm rest}},   
\end{equation} 
where $g$ and $I_{\nu_{\rm rest}}$ are the functions of the light ray, $x^\prime$ and $y^\prime$, 
and we set the observer's distance to be $d=100$ [$M$]. 
As denoted in the previous section, we assume $I_{\nu_{\rm rest}}=B_{\nu_{\rm rest}}$.  
In the same way, for egress phase with the stellar surface at $x$, 
\begin{equation}
L_{\rm bol}(x)
	=\frac{1}{d^2}\int_{-200[M]}^x dx^\prime\int_{-200[M]}^{200[M]}dy^\prime
	\int d\nu_{\rm rest}~g^4~I_{\nu_{\rm rest}}.      
\end{equation} 
We calculate the observed bolometric luminosity 
for the position of the stellar surface $x$ in the range of $-60$[$M$] $<x<60$[$M$]. 

The observed bolometric luminosity can be also the function of time. 
Here, we define the origin of the time when the stellar surface is 
at $60$ [$M$] for the ingress and at $-60$ [$M$] for the egress. 
The region between $\pm 60$ [$M$] are enough to cover the effectively 
emitting region of the accretion disc observed at 1 keV, so we use these 
values in this study when we calculate the normalized light curves. 
In figure \ref{fig:LC}, the upper and lower abscissas are 
the position of the stellar surface $x$ and the time $t$. 
The larger values of the bolometric luminosity is achieved for 
the larger spin parameter of a black hole. 
A rotating black hole transfers angular momentum to the accreting gas 
via the frame-dragging effect which is also calculated when the 
frame transformation by using tetrads. 
This effects increases the observed luminosity. 
Also, the marginally stable circular orbit $r_{ms}$ 
decreases as the black hole spin increases, e.g. $r_{\rm ms}=6M$ 
for $a/M=0$ and $r_{\rm ms}=M$ for $a/M=1$ 
(Bardeen, Press \& Teukolsky 1972). 
Then, the black hole rotation increase the observed luminosity and 
the emission region in the vicinity of the black hole. 
The angular velocity $\omega$ 
of the zero angular momentum observer (ZAMO) with 
respect to distant observer is calculated as 
$\omega=-g_{t\phi}/g_{\phi\phi}=g^{t\phi}/g^{tt}=2Mar/A$. 
Here, $A\equiv (r^2+a^2)^2-a^2\Delta\sin^2\theta$ where $\Delta\equiv 
r^2-2Mr+a^2$. In this study, we assume $\theta=90^\circ$ for the photon 
emitting region. 
Since this angular velocity $\omega$ is roughly proportional to 
$\omega\propto 2Ma/r^3$, 
the effects of the black hole rotation drastically decrease for large $r$. 

\subsection{Normalized Light-Curve Analysis}

\begin{figure}
\includegraphics[width=80mm]{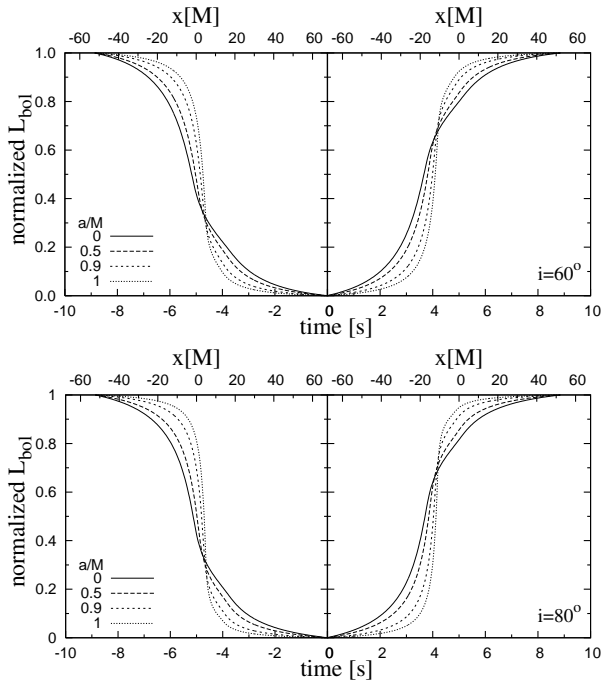}
\caption{
Normalized eclipsed light curves before ({\it left panels}) and after 
({\it right panels}) a total eclipse by a companion stars. 
For each panel, 
the lower abscissa shows the time of eclipse and the upper abscissa 
shows the position of eclipse with respect to the position of 
black hole ($x=0$). 
The ordinate shows the normalized bolometric luminosities defined as 
$L_{\rm bol}(x)/L_{\rm bol}(x=60 M)$ for the ingress and 
$L_{\rm bol}(x)/L_{\rm bol}(x=-60 M)$ for the egress. 
Here, $L_{\rm bol}(x)$ is the observed bolometric luminosity when 
the stellar surface is at $x$. 
That is, for the ingress phase, we normalize the observed bolometric luminosity $L_{\rm bol}(x)$ 
by the observed bolometric luminosity when the stellar surface is at $L_{\rm bol}(x=60M)$, 
and for the egress phase, the observed bolometric luminosity is normalized 
by the observed bolometric luminosity when the stellar surface is at $L_{\rm bol}(x=60M)$. 
In the same way, 
the observed luminosity $L_\nu(x)$ [erg s$^{-1}$ Hz$^{-1}$] can be also 
normalized by the observed luminosity when the stellar surfaces are at $x=\pm 60M$. 
The inclination angles are 
are 60$^\circ$ ({\it top panels}) and 80$^\circ$ ({\it bottom panels}), 
and the black-hole spins are 
$a/M=0$ ({\it solid lines}), $0.5$ ({\it long dashed lines}), 
$0.9$ ({\it short dashed lines}) and $1$ ({\it dotted lines}). 
The lines showing zero time (time$=0$ s) indicate the boundary of a complete eclipse. 
We assumed an orbital velocity of $200$ km s$^{-1}$, a black hole mass of 
$10 M_\odot$ and mass accretion rate of $1 \dot{M}_{\rm Edd}$. 
}
\label{fig:LCspin}
\end{figure}

\begin{figure}
\includegraphics[width=80mm]{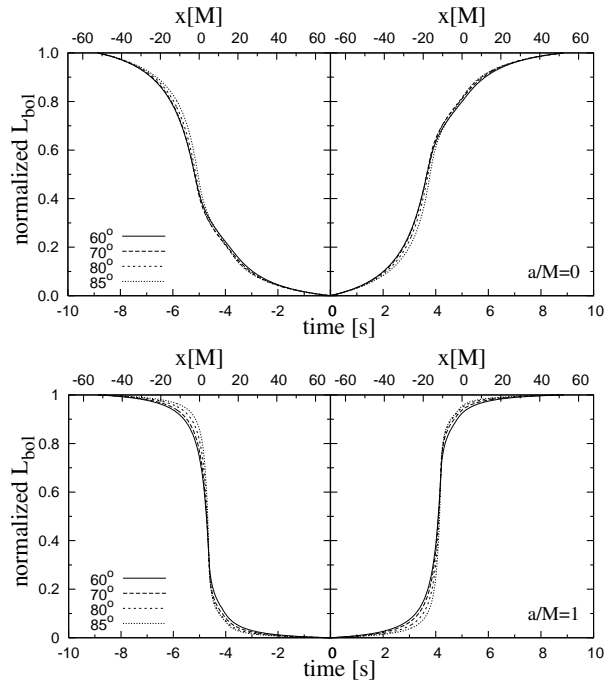}
\caption{
Same as figure \ref{fig:LCspin}. 
The inclination angles are 
are $60^\circ$ ({\it solid lines}), $70^\circ$ ({\it long dashed lines}), 
and $80^\circ$ ({\it short dashed lines}) and 
$85^\circ$ ({\it dotted lines}). 
The black-hole spins are 
$a/M=0$ ({\it top panels}) and $1$ ({\it bottom panels}). 
}
\label{fig:LCincli}
\end{figure}

Although in figure \ref{fig:LC} 
we clearly see the effects of the black hole rotation as the 
increase of the bolometric luminosity, 
it is known that the mass accretion rate also increase the bolometric 
luminosity for some inclination angles (Watarai, Takahashi \& Fukue 2005). 
So, in addition to the maximum bolometric luminosity, 
we analyze the shapes of the light curves which contain the 
information of a black hole spin. 
In order to do this, we define the normalized bolometric luminosity as 
$L_{\rm bol}(x)/L_{\rm bol}(x=\pm 60 M)$. 
That is, for the ingress phase, 
\begin{eqnarray}
\frac{L_{\rm bol}(x)}{L_{\rm bol}(-60)}
	&=&\int_x^{200} dx^\prime\int_{-200}^{200}dy^\prime
	\int d\nu_{\rm rest}~g^4~I_{\nu_{\rm rest}} 
	\nonumber\\	
	&&\bigg/
	\int_{-60}^{200} dx^\prime\int_{-200}^{200}dy^\prime
	\int d\nu_{\rm rest}~g^4~I_{\nu_{\rm rest}}, 
\end{eqnarray} 
where the unit for the length is $M$, and for the egress phase, 
\begin{eqnarray}
\frac{L_{\rm bol}(x)}{L_{\rm bol}(+60)}
	&=&\int_{-200}^x dx^\prime\int_{-200}^{200}dy^\prime
	\int d\nu_{\rm rest}~g^4~I_{\nu_{\rm rest}} 
	\nonumber\\	
	&&\bigg/
	\int_{-200}^{60} dx^\prime\int_{-200}^{200}dy^\prime
	\int d\nu_{\rm rest}~g^4~I_{\nu_{\rm rest}}. 
\end{eqnarray} 

Since we calculate the eclipsing light curves in the region of 
$-60M<r<60M$ in the present study, 
$L_{\rm bol}(x=\mp 60M)$ represent the maximum bolometric luminosity 
at the ingress and the egress, respectively. 
Figure $\ref{fig:LCspin}$ shows the calculated normalized light curves 
at the ingress ({\it left panel}) and egress ({\it right panel}) 
of an eclipse. 
From this figures, we see that the black hole rotation clearly deforms 
the shapes of the normalized light curves. 
For larger spin parameters, the light curves changes in the closer 
region of the black hole center ($x=0$[$M$]). 
This is because the effective emission region of photon from 
the accretion disc become more compact for larger spin parameter 
as shown in figure \ref{fig:BHimage}. 
Then, the drastically changed timescale of the normalized luminosity 
become shorter for the larger spin parameter. 
We also investigate the effects of the inclination angle $i$. 
At first glance of the upper panels and the lower panels of 
figure \ref{fig:LCspin}, there is little dependence on the inclination 
angle. 
In order to see the effects of the inclination angles on the shapes of the 
light curves, in figure \ref{fig:LCincli} 
we plot the normalized light curves for the black hole spin of $a/M=0$ 
({\it upper panels}) and $1$ ({\it lower panels}). 
Although the absolute value of the bolometric luminosity depends on the 
inclination angle largely as we can see in figure \ref{fig:LC},  
from figure \ref{fig:LCincli} the effects of the inclination angle on the 
normalized light curves is very small. 
Then, the shape and the variation timescale of 
the normalized light curve are the possible candidates 
of the good indicator of the black hole spin. 

\subsection{Skewness and Kurtosis}

In addition to the normalized light curves given in the previous section, 
we present the other indicator of the effects of the black hole spin. 
We introduce the statistical quantities of skewness, $S$, and kurtosis, 
$K$, as indicators of the light-curve asymmetry. 
These are the same as the analysis in Watarai, Takahashi \& Fukue (2005). 
The skewness and kurtosis represent the deviation of the observational 
data from the Gaussian (normal) distribution. 
Skewness, $S$, is a measure of the degree of asymmetry of a distribution. 
On the other hand, kurtosis, $K$, is the degree of peaky feature of 
a distribution. 

We have done the skewness and kurtosis analysis for the inverted 
bolometric light curves as calculated in figure \ref{fig:LC}. 
From the data of the inverted normalized light curve, we first obtain 
$f_i$ as $f_i=L_{\rm bol, max}-L_{\rm bol}(t_i)$ where $t_i$ is the time 
of each mesh. 
The statistical quantities are defined as  
\begin{eqnarray}
P_i&=&f_i/\sum_{i=0}^n f_i,\\ 
\bar{t}&=&\sum_{i=0}^n t_i P_i,\\ 
\sigma^2&=&\sum_{i=0}^n (t_i-\bar{t})^2 P_i,\\ 
S&=&\frac{1}{\sigma^3}\sum_{i=0}^n (t_i-\bar{t})^3 P_i,\label{eq:S}\\   
K&=&\frac{1}{\sigma^4}\sum_{i=0}^n (t_i-\bar{t})^4 P_i,\label{eq:K}\\
\end{eqnarray}
where $n$ is the number of data points, 
$P_i$ is the normalized distribution calculated from the bolometric 
light curves as shown in figure \ref{fig:LC}, 
$\bar{t}$ is an weighted average time 
and $\sigma$ is the standard deviation. 
If the left tail is more pronounced than the right tail, 
the skewness becomes negative ($S<0$).  
If the reverse is true, the skewness becomes positive ($S>0$). 
The skewness for a normal distribution is zero $S=0$, 
and any symmetric data should have a skewness near zero. 
The kurtosis for a standard normal distribution is three. 
Positive kurtosis indicates a peaked distribution and negative 
kurtosis indicates a flat distribution. 

\begin{figure*}
\includegraphics[width=170mm]{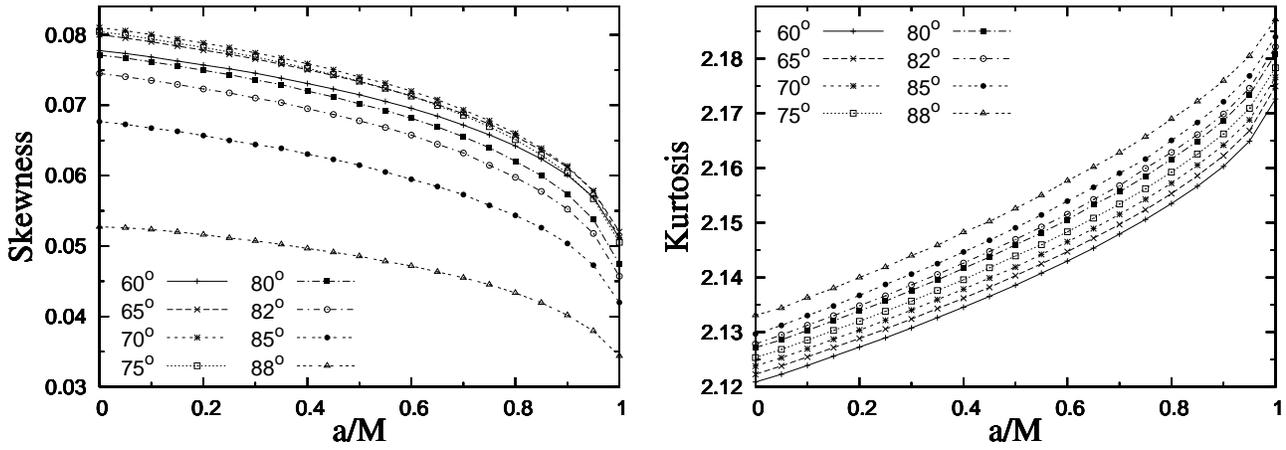}
\caption{
Skewness and kurtosis for different black-hole spins and inclination
angles, $i=60^\circ$, $65^\circ$, $70^\circ$, $75^\circ$, $80^\circ$, 
$82^\circ$, $85^\circ$ and $88^\circ$. 
}
\label{fig:SK}
\end{figure*}

In figure \ref{fig:SK}, we show the dependence of 
the calculated skewness, $S$ ({\it left panel}) 
and kurtosis, $K$ ({\it right panel}) on the black hole spin, $a/M$, 
for eight inclination angles ($60^\circ$, $65^\circ$, $70^\circ$, 
$75^\circ$, $80^\circ$, $82^\circ$, $85^\circ$ and $88^\circ$). 
Since the marginally stable orbit become larger for non-rotating 
black hole than the rotating cases, then the emission region 
become wider than the rotating one. 
Then, for larger spin parameters, the skewness becomes smaller value. 
This is clearly seen in the left panel of figure \ref{fig:SK}. 
On the other hand, for larger spin parameters, the emission region and 
the variation timescale of the light curves become smaller. 
Then the light curves are expected to be more peaky than cases of 
the non-rotating black hole. 
Then, the kurtosis become larger value for the larger spin parameters of 
black holes. 
This feature can be seen in the right panel of figure \ref{fig:SK}. 
While the kurtosis cannot be good indicator of the mass accretion rate 
for non-rotating black holes as shown in 
Watarai, Takahashi \& Fukue (2005), 
both of the skewness and the kurtosis can be the possible indicator of
the black hole spin. 

\section{Eclipsing Light Curves with Atmospheric Effects of Companion Star} 

While in the last section we do not take into account the atmospheric 
effects of the companion star, 
in this section, we calculate the eclipsing light curves with the atmospheric effects. 
One of the key questions in terms of the atmospheric effects 
are whether the information of the black hole spin 
in the eclipsing light curves 
is completely smeared out by the atmospheric effects.  
One of the main goals in this section is to answer this question. 
We also investigate how the eclipsing light curves depend on 
the observed photon energy in X-ray and the inclination angle.  

\subsection{Black Hole Shadows with the Atmospheric Effects}
\begin{figure*}
\includegraphics[width=170mm]{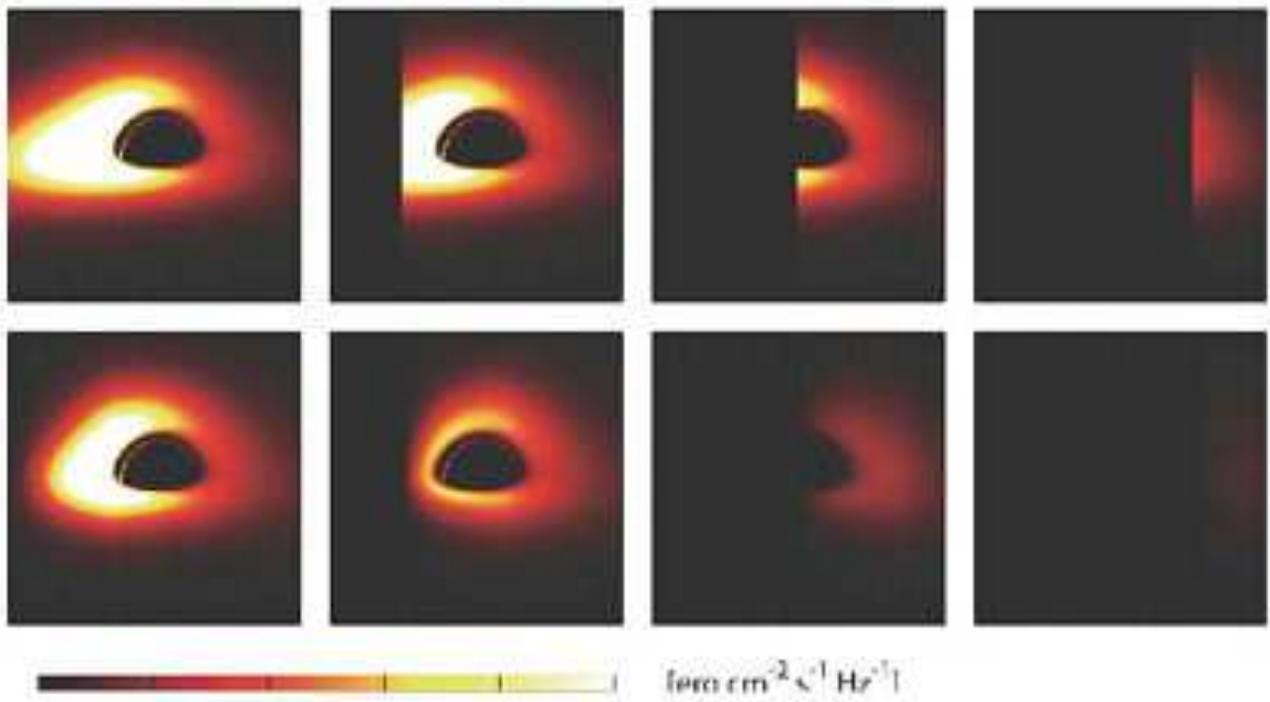}
\caption{
Images of black hole shadows with $a/M=0.5$ when ingress 
at the observed frequency of 1 keV for cases with no absorption 
effects ({\it top panels}) and atmospheric absorptions 
({\it bottom panels}).  
The size of each image is same as figure \ref{fig:BHimage}, i.e. 
$-20<x/M<20$ and $-20<y/M<20$. 
For no absorption case for top panels, 
the edges of the companion star are located at 
-20, -10, 0 and 10 in units of $M$ ({\it left to right}). 
For the absorption case for bottom panels, we define these edges 
as positions with $\tau(\nu)=7$. 
}
\label{fig:BHabs}
\end{figure*}

%
We first calculate the images of the black hole shadows with absorption 
effects in the atmosphere of the companion star. 
In figure \ref{fig:BHabs}, we show 
images of black hole shadows when ingress 
at the observed frequency of 1 keV for cases with no absorption 
effects ({\it top panels}) and atmospheric photoionization absorptions 
({\it bottom panels}). 
We assume the mass of the companion star to be 1 $M_\odot$.  
The size of each image is same as figure \ref{fig:BHimage}, i.e. 
-20 [$M$]$<x<$20 [$M$] and -20 [$M$]$<y<$20 [$M$]. 
For no absorption case for top panels, 
the edges of the companion star are located at 
-20, -10, 0 and 10 in units of $M$ ({\it left to right}). 
For the cases with the atmospheric effects ({\it bottom panels}),  
we define these edges 
as positions with $\tau(\nu)=7$. 
We can see that the brightness around the black hole shadow are 
decreased by the photoionization absorption in the atmosphere. 
In addition to the decrease of the absolute values of the brightness, 
the clear edges of the companion star which can be seen for no absorption 
cases are smeared out by the absorption effects. 
This is because the effective region with absorption effects is 
comparable to the size of X-ray emitting regions. 

\subsection{Normalized Light Curves with the Atmospheric Effects}
\begin{figure}
\includegraphics[width=80mm]{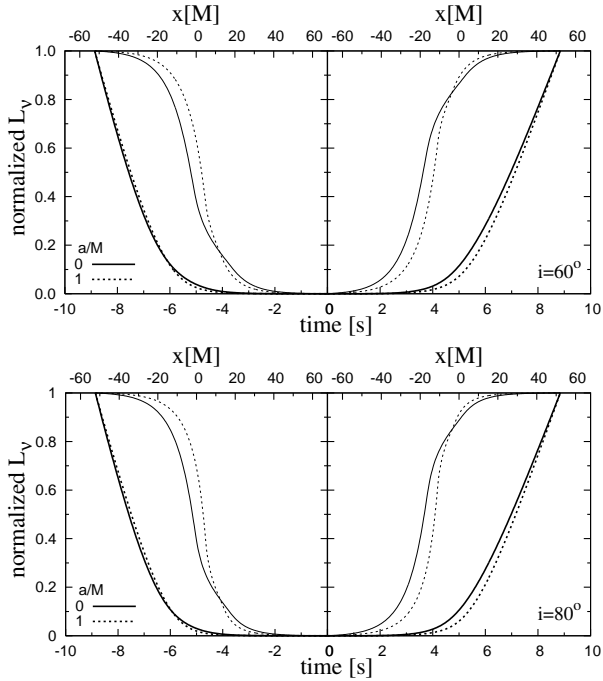}
\caption{
Normalized light curves 
with ({\it thick lines})
and 
without ({\it thin lines})
the atmospheric effects of the companion star 
at observed energy of 1 keV for 
black hole spins $a/M=0$ ({\it solid lines}) and 1 ({\it dotted lines}), 
and the observed inclination angles with respect to the rotation axis 
of the accretion disc $i=60^\circ$ ({\it top panels}) and 
$80^\circ$ ({\it bottom panels}).  
}
\label{fig:norLCabs_1keV}
\end{figure}

We next calculate the eclipsing light curves with effects of the 
photoionization absorption. 
In figure \ref{fig:norLCabs_1keV}, we show 
normalized light curves 
with ({\it thick lines})
and 
without ({\it thin lines})
the atmospheric effects of the companion star.
The observed energy is set to be 1 keV in figure \ref{fig:norLCabs_1keV}. 
The black hole spins are $a/M=0$ ({\it solid lines}) and 1 ({\it dotted lines}), 
and the observed inclination angles with respect to the rotation axis 
of the accretion disc are $i=60^\circ$ ({\it top panels}) and 
$80^\circ$ ({\it bottom panels}). 
While the normalization light curves with no absorption effects 
largely change in a short time when the most luminous parts of the 
accretion disc are occulted or unocculted by the sharp edge of 
the companion star, 
the normalization light curves with absorption effects 
gradually decrease or increase compared to the cases with no absorption 
effects. 
This is because the atmosphere of the companion star absorb some fraction 
of X-ray photons gradually before the most luminous part of the accretion 
disc is completely occulted or unocculted. 
In other words, the width of the absorption region which causes the 
gradual changes of the X-ray brightness of the accretion disc 
is comparable to size of the luminous parts of the accretion disc as 
denoted above. 
The differences between 
the normalized light curves with absorption effects for $a/M=0$ and 1 
become smaller than the cases of no absorption effects. 
When egress, i.e. $t>0$, the normalized light curves with $a/M=0$ 
rise more rapidly than the cases of $a/M=1$. 
That is to say, the timescale of the increase of normalized $L_\nu$ 
for $a/M=0$ is longer than that for $a/M$. 
This is because the most luminous parts of the accretion disc for 
$a/M=1$ is more compact than those for $a/M$. 
This results in the shorter crossing time of the companion star along 
the luminous parts of the disc for $a/M=1$ than the cases for $a/M=0$. 
This feature can be seen in the case of no absorption by 
the companion star. 
Since the size of the most luminous parts of the accretion disc 
depends on the black hole spin, 
the crossing timescale of the companion star along the most luminous parts 
of the accretion disc can be one of the key physical quantities 
which distinguish the cases of rapidly rotating black holes and 
the cases of no rotating black hole. 
But these effects are generally 
smeared out by the atmospheric effects of the companion star 
at some fraction. 
In our calculations assuming the solar atmosphere, 
the atmospheric absorption can not completely smeared out the difference 
between the cases of rapidly rotating black holes and 
the cases of no rotating black hole. 
Then, the observed eclipsing light curves at 1 keV with the 
atmospheric effects of the companion star possibly contain 
the information of the black hole spin. 
However, since the effects of the atmosphere is larger than the effects of the 
spin, only when we have the reliable atmospheric model, 
we can determine the black hole spin from the eclipsing light curves 
observed by the wavelength where the atmospheric absorption is effective. 
%

\subsection{Observed Photon Energy}

\begin{figure*}
\includegraphics[width=170mm]{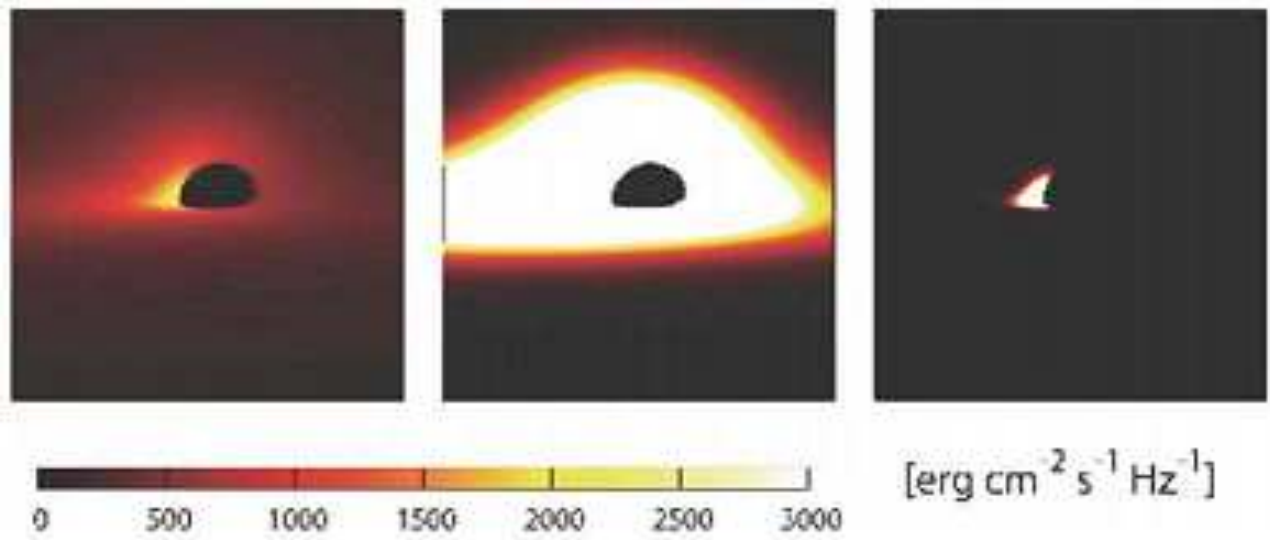}
\caption{
Images of the black hole shadows in the accretion discs for 
observed energy of 
0.1 keV ({\it left panel}), 
1 keV ({\it middle panel}) and 
10 keV ({\it right panel}). 
In these panels, the inclination angle of the observer is $i=80^\circ$ and 
the black hole spin is $a/M=1$. 
}
\label{fig:BHkeV_a10}
\end{figure*}

Here, we investigate the observed energy dependence of the atmospheric 
absorption effects in the images of the black hole shadows 
and the normalized light curves. 
In figure \ref{fig:BHkeV_a10}, 
images of the black hole shadows in the accretion discs 
are shown for observed photon energy of 
0.1 keV ({\it left panel}), 
1 keV ({\it middle panel}) and 
10 keV ({\it right panel}). 
Here, we do not include the atmospheric absorption effects. 
In these panels, the inclination angle of the observer is $i=80^\circ$ and 
the black hole spin is $a/M=1$. 
The absolute values of the luminosities are largest for the case 
of 1 keV. 
On the other hand, 
the larger the observed photon energy becomes, 
the more compact 
the regions of the luminous parts in the accretion discs become. 
Especially, 
in the case of 10 keV, 
the size of the luminous parts in the accretion disc is smallest and 
the luminous parts exist in the blue shifted parts of the 
accretion disc which are located in just outside the black hole shadow. 
As Takahashi (2004) calculated, 
the size and the positions of the black hole shadows in the accretion 
discs clearly depend on the black hole spins. 
Thus, the eclipsing light curves for higher photon energy can 
contain the information of the black hole shadows in case that 
the atmospheric smearing effects are negligible.

\begin{figure}
\includegraphics[width=80mm]{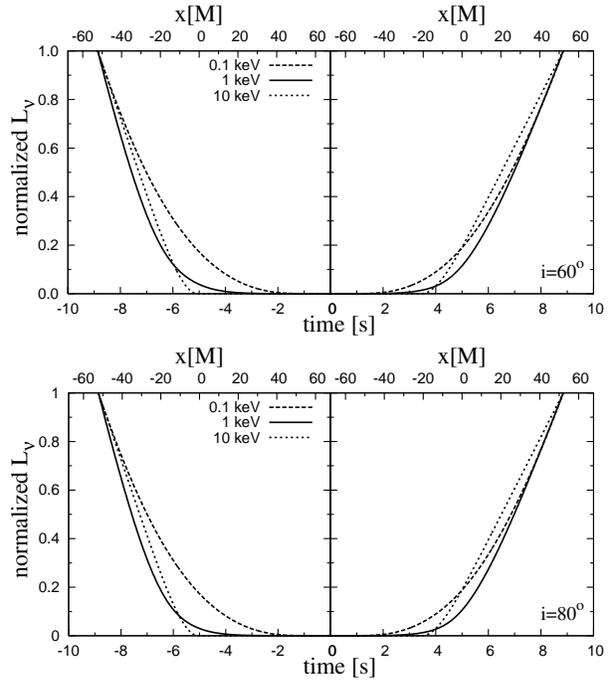}
\caption{
The normalized light curves with effects of the atmospheric absorption 
for observed energy of 
0.1 keV ({\it dashed lines}), 
1 keV ({\it solid lines}) and 
10 keV ({\it dotted lines}). 
The black hole spin is $a/M=0$ and 
the inclination angles of the observers are 
$i=60^\circ$ ({\it top panel}) and 
$80^\circ$ ({\it bottom panel}). 
}
\label{fig:norLCabs_keV_a00}
\end{figure}

\begin{figure}
\includegraphics[width=80mm]{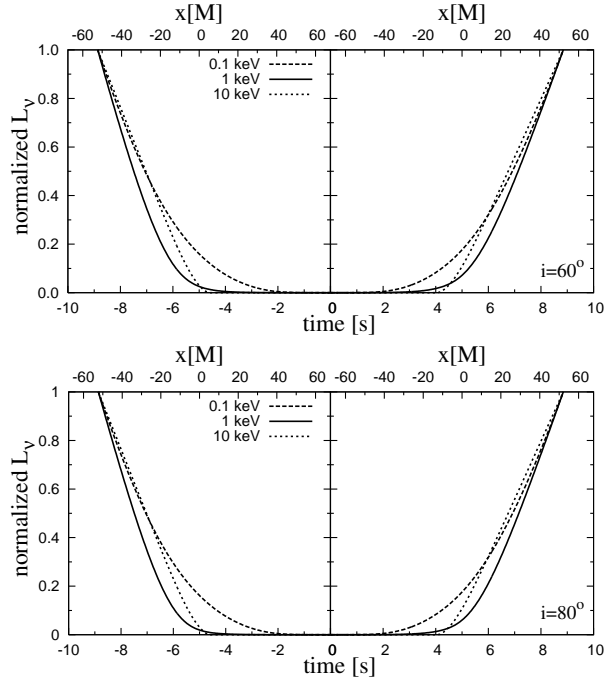}
\caption{
The same as figure \ref{fig:norLCabs_keV_a00} but for $a/M=1$. 
}
\label{fig:norLCabs_keV_a10}
\end{figure}

The normalized light curves with effects of the atmospheric absorption 
are shown in figure \ref{fig:norLCabs_keV_a00} when $a/M=0$ 
and in figure \ref{fig:norLCabs_keV_a10} when $a/M=1$ 
for observed energy of 
0.1 keV ({\it dashed lines}), 
1 keV ({\it solid lines}) and 
10 keV ({\it dotted lines}). 
The black hole spin is $a/M=0$ and 
the inclination angles of the observers are 
$i=60^\circ$ ({\it top panels}) and 
$80^\circ$ ({\it bottom panels}). 
The differences between cases of 0.1 keV, 1 keV and 10 keV are 
clearly shown in these figures. 
The higher the observed photon energy become, 
the shorter the timescales of the changes of the normalization light curves 
become. 
This is because the larger the observed photon energy becomes, 
the more compact 
the regions of the luminous parts in the accretion discs become and then 
the crossing time of the companion star is shortest in the case of 10 keV 
and longest in the case of 0.1 keV. 
These features can be seen in both cases of 
$a/M=0$ in figure \ref{fig:norLCabs_keV_a00} 
and $1$ in figure \ref{fig:norLCabs_keV_a10}.

\subsection{Black Hole Spin}

\begin{figure}
\includegraphics[width=80mm]{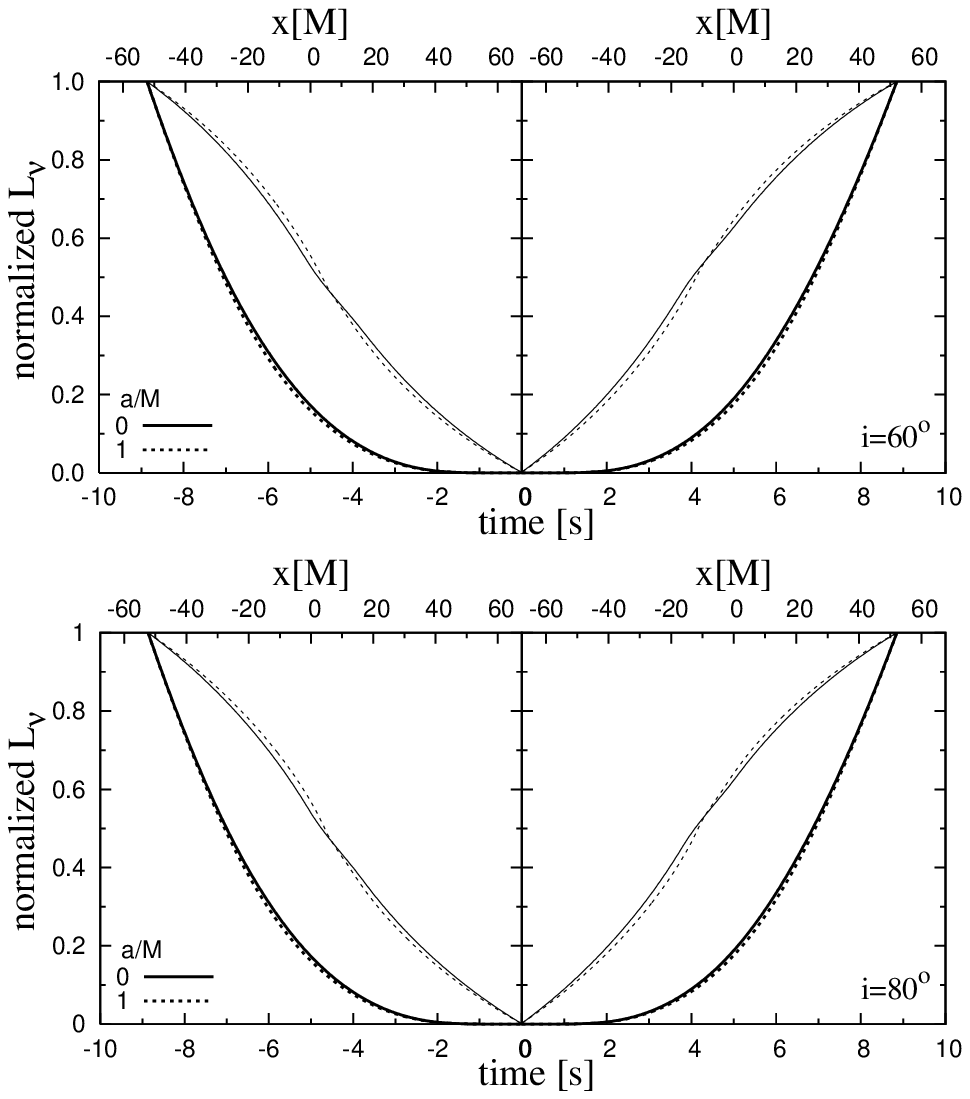}
\caption{
The same as figure \ref{fig:norLCabs_1keV} but for 
the observed photon energy of 0.1 keV. 
}
\label{fig:norLCabs_01keV}
\end{figure}

\begin{figure}
\includegraphics[width=80mm]{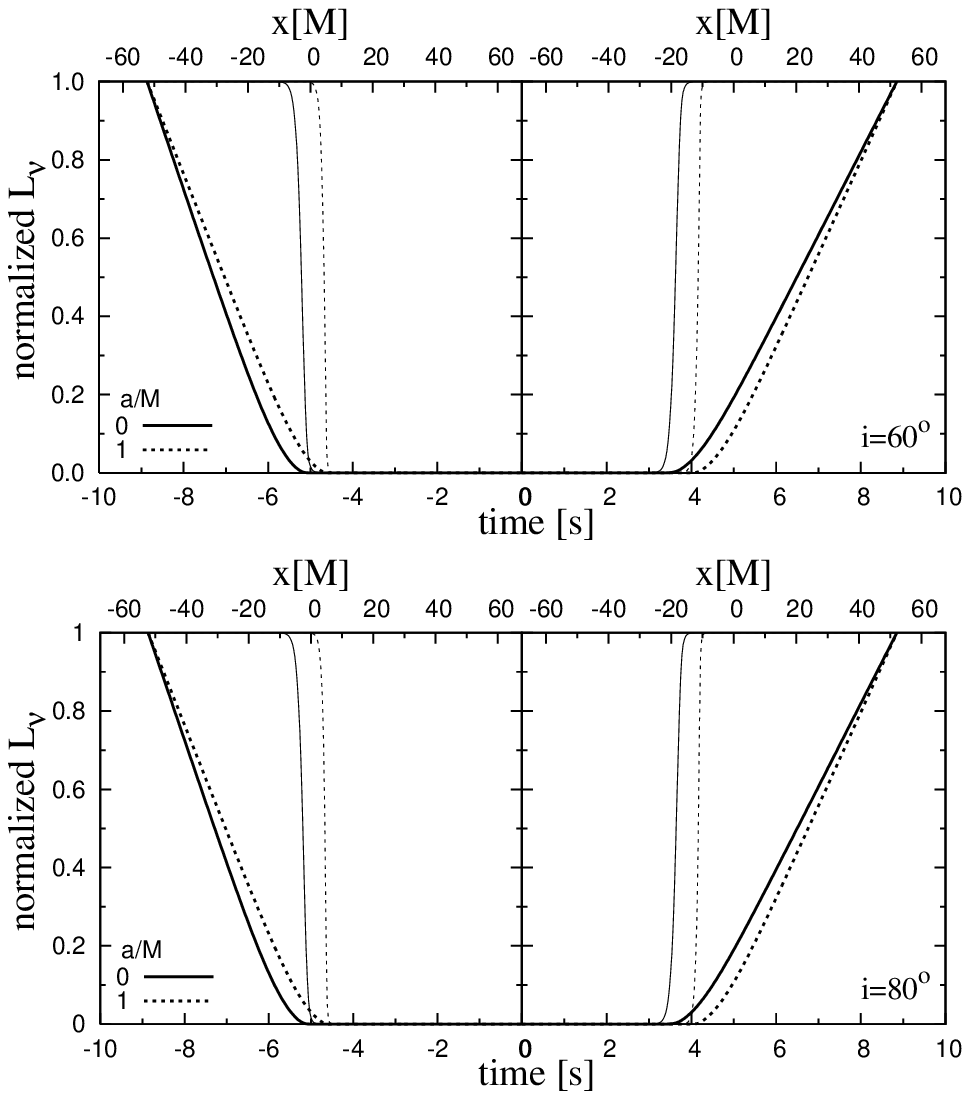}
\caption{
The same as figure \ref{fig:norLCabs_1keV} but for 
the observed photon energy of 10 keV. 
}
\label{fig:norLCabs_10keV}
\end{figure}

In order to see the effects of the black hole spin in the normalized 
light curves when the observed photon energies are 0.1 keV and 10 keV, 
we plot the normalized light curves 
same as figure \ref{fig:norLCabs_1keV} but for 
the observed photon energy of 0.1 keV in figure \ref{fig:norLCabs_01keV} 
and 10 keV in figure \ref{fig:norLCabs_10keV}. 
In the case of 0.1 keV in figure \ref{fig:norLCabs_01keV}, 
the difference of the normalized light curves 
between the cases of $a/M=0$ and the cases of $a/M=1$ 
are very small for both cases of $i=60^\circ$ and $80^\circ$. 
Thus, it seems very difficult to see the information of the black hole spin 
in the observed light curves. 
On the other hand, in the case of 10 keV, 
we can clearly see the difference of the normalized light curves 
between the cases of $a/M=0$ and the cases of $a/M=1$ 
for both cases of $i=60^\circ$ and $80^\circ$. 
The timescale of the change of the normalized light curve for the 
case of $a/M=1$ is longer than the timescale for the case of $a/M=0$ 
for both cases of ingress and egress. 
The normalized light curves are gradually decrease and increase 
because of the atmospheric effects and 
the most luminous 
parts of the accretion discs exist in just outside of the black hole 
shadows as denoted above. 
Moreover, the size of the black hole shadows become smaller 
for larger spin and the position of the black hole shadow are shifted to 
the opposite side of the most luminous parts of the accretion discs
(Takahashi 2004). 
The timescale when the most luminous parts can be seen by 
the observer become longer when $a/M=1$ than the case of $a/M=0$ because 
the most luminous parts exist in more inner regions when $a/M=1$ than the 
case of $a/M=0$. 

\subsection{Inclination Angle between the Rotation Axis and the Direction of 
the Observer}
\begin{figure}
\includegraphics[width=80mm]{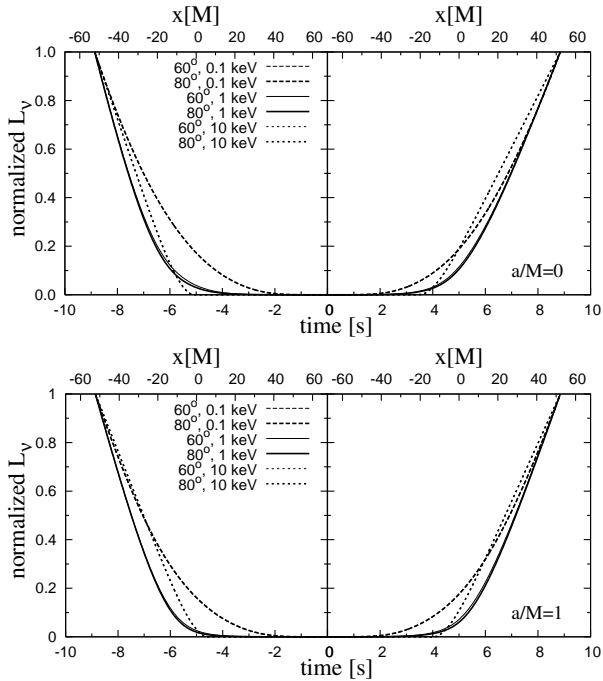}
\caption{
The normalized light curves for cases of $a/M=0$ ({\it top panel}) 
and $1$ ({\it bottom panel}). 
For each panel, the lines are plotted for the combinations of 
the inclination angle $i=60^\circ$, $70^\circ$ and $80^\circ$ 
and the observed photon energy to be 0.1 keV, 1 keV and 10 keV. 
}
\label{fig:norLCabs_keV_incli}
\end{figure}

We also investigate the dependence of the normalized light curves 
on the observed inclination angles, $i$, 
with respect to the rotation axis of the accretion disc. 
While so far we plot the normalized light curves for both cases of 
$i=60^\circ$ and $80^\circ$, 
in order to see the dependence on the inclination angle $i$, 
we plot the normalized light curves with same black hole spins 
in the same panel. 
The normalized light curves for cases of $a/M=0$ ({\it top panel}) 
and $1$ ({\it bottom panel}) are shown in figure 
\ref{fig:norLCabs_keV_incli}. 
For each panel, the lines are plotted for the combinations of 
the inclination angle $i=60^\circ$, $70^\circ$ and $80^\circ$ 
and the observed photon energy to be 0.1 keV, 1 keV and 10 keV. 
While the observed images of the black hole shadows in the accretion 
discs are clearly different for the different inclination angles $i$,  
the dependences on the inclination angles are very small in the 
normalized light curves. 

\section{Discussion}

\subsection{Eclipsing Light Curves and Observational Feasibility}
Recently, Pietsch et al. (2006) give the clear observational data of the 
eclipsing light curves for the X-ray binary M33 X-7 
detected by {\it Chandra}. 
While the absorption effects in the companion atmosphere seem to vanish the 
relativistic effects in the eclipsing light curves 
in their observational data, 
our calculations show that the information of the black hole spin in the 
eclipsing light curves can not be completely smeared out for the observed 
photon energy of 1 keV and 10 keV if we assume the solar-type 
atmospheric model. 
However, as denoted above, even if we have the reliable observational data 
with high signal-to-noise ratio, 
only when we have the reliable atmospheric model of the companion star, 
the physical information of the strong-gravity regime, 
such as, a black hole spin, can be obtained. 

In addition, in the present study we use the solar atmospheric model 
as a basic sample of the atmospheric structure. 
In the case of the eclipsing X-ray binary M33 X-7, 
the companion star is an O6III star. 
Although some recent observational data suggest the existence of corona 
in the atmosphere of Herbig Ae/Be stars (Stelzer et al. 2006), 
it remains one of the unsolved issues whether 
the high temperature stars such as O-type stars 
have corona in their atmosphere. 
If there is the corona in the atmosphere of O-type stars, 
the stellar atmosphere will have two-temperature structure such as the 
solar atmosphere. 
In such cases, our calculations roughly calculate the light curves for 
the case of M33 X-7. 

Basically, the count rates of the X-ray photons 
from X-ray binaries in external galaxies 
with the possible exception of the Magellanic Clouds are generally 
much too low to provide the precise eclipse mapping of the inner 
portions of the accretion discs even in Chandra observations. 
In addition, the time resolutions for the data of the eclipsing light curves 
for M33 X-7 are insufficient to precisely map the small strong-gravity regime 
in the center of the accretion flows. 
Even in the case that the calculations in the present study 
are not used to the present low count rate observations in X-ray, 
future X-ray data with higher count rates obtained by the future 
X-ray telescope such as Constellation-X Observatory in NASA 
(see, http://constellation.gsfc.nasa.gov/) or 
XEUS (X-Ray Evolving Universe Spectrometer) in ESA 
(see, http://www.esa.int/science/xeus) 
will provide the information of the extremely strong-gravity regions 
of the accretion discs. 

\subsection{Accretion Disc Model}
While our calculation assumes the standard accretion disc, 
for the accretion discs with larger mass accretion rate, 
Watarai, Takahashi \& Fukue (2005) pointed out the light curves are 
deformed by the self-occultation effects, and shows that 
for larger mass accretion rate the skewness becomes small value and 
the kurtosis is nearly constant. 
On the other hand, our analysis shows the larger the black hole spin becomes, 
the smaller the skewness becomes and the larger the kurtosis becomes 
for any observed inclination angles.  
The dependence of the kurtosis on the black hole spins contrast with 
the dependence of the mass accretion rate. 
We expect that the effects due to the black hole rotation can not be seen 
for discs with large mass accretion rate because of the self-occultation 
effect. 
The same analysis for the accretion discs with high mass accretion rate 
around rotating black hole is a topic for future study. 
Moreover, 
from the eclipsing light curves, it may be possible to examine 
the disk structure such as stress at the marginally stable orbit, 
the angular dependence of the emitted radiation, and the dynamical 
structure relating the magnetic processes such as magnetorotational instabilities 
in the supersonic region which are the fundamental unsolved issues about 
the accretion flow models.  
These issues are also topics for future studies. 

\subsection{X-ray Heating Effects}
So far, we implicitly neglect the effects of the heating of the atmosphere 
due to the absorbed X-ray photons in the atmosphere of the companion star. 
According to the calculations of the structure of the heated atmosphere 
(e.g. Basko \& Sunyaev 1973), the temperature of the atmosphere 
increase by about factor 2 due to the absorbed X-ray photons. 
The number density is also changed by similar factor.  
Although these changes causes the quantitative changes of the optical 
depth in X-ray of the atmosphere of the companion star, 
we do not expect the qualitative changes of the dependence of the 
eclipsing light curves on the physical parameters. 
Since the absorption effects basically do not change the timescale of the 
change of the eclipsing light curves, 
the effects of the X-ray heating will not drastically alter the dependence 
of the crossing timescale of the luminous parts of the accretion disc. 
In addition to the changes of the optical depths, 
a X-ray heating effects causes 
a radiatively driven stellar wind originated from 
the X-ray heated atmosphere (e.g. Blondin 1994, Wojdowski et al. 2001). 
The atmospheric profile of the companion star is not the same 
between ingress and egress due to this effect. 
Since this effects also change the optical depth of the atmosphere in 
X-ray, we expect the effect will not change 
the timescale of the change of the eclipsing light curves drastically.

\subsection{Cases for Neutron Stars and Seyfert Galaxies}
Asymmetric brightness distribution exists not only in black hole X-ray
binaries, but also in neutron star X-ray binaries (NSXBs). 
This is because the relativistic effects still works in NSXBs. 
Even in the case that the emission from the neutron star is dominant, 
the asymmetric component still remains in its light curves
as long as the emission from the neutron star surface is isotropic.
The observational data for the eclipsing neutron star binaries are actually 
obtained (e.g. Homan et al. 2003). 
In the case of optically thin along the light paths,  
we can estimate the emitting radius from the duration time
of ingress or egress, 
and measure the spin of the neutron star from the skewness analysis 
of the asymmetric eclipsing light curves. 
According to our results,
the result does not depend on the inclination angle. 
Weaver \& Yaqoob (1998) shows the "deep minimum" in the X-ray light curves 
of Seyfert 1 galaxy MGC-6-30-15. 
Some obscuring body in the line of sight seem to 
cause the occultation phenomena in the light curves. 
Only when such events occur many times and the typical features of the 
light curves are obtained, and the photons from the vicinity of the black hole 
are directly detected, 
it may be possible to obtain the physical 
information in the vicinity of the black hole, 
such as the size of emission region and black hole spin.  

\section{Concluding Remarks}
In this paper, we propose an eclipsing light-curve diagnosis 
for accretion flows around rotating black holes. 
When emission from an inner part of the accretion disc around black hole 
is occulted by a companion star, 
the light curves at ingress and egress show the asymmetric features 
due to the relativistic effects such as Doppler boosting originally 
pointed out by Fukue (1987). 
Our calculations include the asymmetric feature due to the black hole 
rotation. 
After we give the calculation method of the eclipsing light curves 
with and without the atmospheric effects of the companion star in 
\S 2.1 and \S 2.2, respectively, 
the numerical calculation of the eclipsing light curves are 
performed and the analysis for the curves are given. 
When no atmospheric effects of the companion star, 
the effects of the black hole rotations on the light curves are 
clearly seen in 
the light curves (\S 3.1), 
the normalized light curves (\S 3.2) 
and the statistical quantities of skewness and kurtosis (\S 3.3). 
We show that 
the eclipsing light curves without the atmospheric effects clearly 
reflect the effects of the black hole's rotation. 
We also investigate the eclipsing light curves with the atmospheric 
effects of the companion star. 
Based on the adopted atmospheric model, after evaluating the absorption and 
scattering effects in X-ray, we calculate the images of 
the black hole shadows in \S 4.1, 
the normalized light curves \S 4.2. 
The analysis of the eclipsing light curves with the atmospheric effects 
are given in terms of 
the observed photon energy (\S 4.3), the black hole spin (\S 4.4) and 
the inclination angle (\S 4.5). 
The higher the observed photon energy become, 
the shorter the timescales of the changes of the normalization 
light curves become. 
This is because  
the crossing time of the companion star is shortest in the case of 10 keV 
and longest in the case of 0.1 keV. 
Since the size of the most luminous parts of the accretion disc 
depends on the black hole spin, 
the crossing timescale of the companion star along the most luminous parts 
of the accretion disc can be one of the key physical quantities 
which distinguish the cases of rapidly rotating black holes and 
the cases of no rotating black hole. 
In our calculations assuming the solar atmosphere, 
while the effects of black hole spins are negligible 
when the observed photon energy of 0.1 keV 
for both cases 
with no atmospheric effects and with atmospheric effects, 
for the observed photon energies of 1 keV and 10 keV 
the atmospheric absorption can not completely smeared out the difference 
between the cases of rapidly rotating black holes and 
the cases of no rotating black hole. 
However, these effects are 
smeared out by the atmospheric effects of the companion star 
at some fraction, and  
in our atmospheric model, the effects of the atmosphere are 
much larger than the effects of the black-hole spin. 
Therefore, even in the case that the light-curves contain the information 
of the black hole spin, it may be difficult to extract the information of the 
black hole spin if we do not have the realistic atmospheric profiles, such as, 
the temperature, the number densities for several elements. 
Even in such cases, based on our calculations, 
the light curve asymmetries due to the rotation of the accretion disc exist. 
Only when we have the reliable atmospheric model, in principle, 
the black hole spin can be determined 
from the eclipsing light curves observed around 1 keV.
%




\section*{Acknowledgments}

One of the author (RT) express his thanks to T. Suzuki, M. Shibata, 
I. Hachisu and Y. Sekiguchi for valuable comments and discussion.  
The author (RT) is grateful to Y. Eriguchi and S. Mineshige 
for their continuous encouragements. 
This research was partially supported by the Ministry of Education, Culture, 
Sports, Science and Technology, Grant-in-Aid for 
Japan Society for the Promotion of Science 
(JSPS) Fellows (17010519 RT; 16004706 KW).



\bsp

\label{lastpage}

\end{document}